\documentstyle[12pt]{article}

\begin{document}

\title{ Multiparticle Threshold Amplitudes Exponentiate in
Arbitrary Scalar Theories}
\author{ M.V.Libanov \\ {\small {\em Institute
  for Nuclear Research of the Russian Academy of Sciences,}}\\
  {\small {\em 60th October Anniversary prospect, 7a, Moscow
  117312, Russia}}\\
  {\small {\em and}}\\
  {\small {\em Physics Department, Moscow State University, Moscow,
  Russia.}}}
\date{October, 1996} \maketitle
\begin{abstract}
Threshold amplitudes are considered for
$n$-particle production in arbitrary scalar theory. It is found
that, like in $\phi ^4$, leading-$n$ corrections to the tree level
amplitudes, being summed over all loops, exponentiate. This result
  provides more evidence in favor of
the conjecture on the exponential behavior of the
multiparticle amplitudes.
\end{abstract}

\newpage
\paragraph{1.} Considerable interest has been attracted in recent years
to the issue of multiparticle production both in perturbative and
non-perturbative regimes in weakly coupled scalar theories (for
reviews, see\,\cite{VolRoch,ECH}). This problem has been initiated
by the qualitative observation\,\cite{Corn,Gold} that in the
 $\lambda \phi ^4$ theory the amplitudes of the processes
of creation of a large number of bosons by a few initial particles
exhibit factorial dependence on the multiplicity of the final
state. The reason is that the number of tree graphs contributing to
the amplitude of creation of $n$ particles grows as $n!$. At $n\sim
1/\lambda $ this factor is sufficient to compensate the suppression
due to the small coupling constant, and tree level amplitudes
become large.

By now, numerous perturbative
results in different scalar models have been obtained that
confirm the factorial growth of the tree level amplitudes (see,
e.g., refs.\,\cite{Voloshin,Argyres,Brown,BrownZ,Smith,LRST,LRT})
and, what is more important, suggest the
exponential behavior of multiparticle cross sections~\cite{LRST},
\begin{equation}
\sigma (E,n)\sim\exp\left( \frac{1}{\lambda }F(\lambda
n,\epsilon ) \right)
\label{2}
\end{equation}
where $\epsilon =(E-nm)/(nm)$ is the typical kinetic energy of
outgoing particles, $E$ is the energy of an initial state, and
$\lambda $ is the typical  coupling constant of the
theory. Moreover, there exist several results\,\cite{????}
on the first loop corrections at threshold to the tree level
amplitudes $A_{\rm tree}$.  It  turns out that  these
corrections are of order  $\lambda n^2A_{{\rm tree}}$ and are
comparable with the tree level contributions even at $\lambda  n
< 1$. So, to obtain the correct expressions for the
multiparticle amplitudes, one has to take into account all
loops.  Presently there are  two
realistic approaches to calculate all leading loop contributions.
The first approach has been suggested in models with $O(N)$ symmetry
and in models with softly broken $O(N)$ symmetry in the regime
$N\to\infty$ \cite{Makeenko,Makeenkoetal}. The result coincides
with the tree level one with all parameters (masses and coupling
constants) replaced by their renormalized values.  The second
approach has been developed in the context of $\lambda \phi ^4$
theory (both with broken and unbroken reflection symmetry) for
various initial states\,\cite{LRST,LST} with the result that the
threshold amplitude is
\begin{equation}
A_{{\rm all-loop}}(\epsilon
=0)=A_{{\rm tree}}{\rm e}^{B\lambda n^2+O(\lambda n,\,\lambda
^2n^3)}
\label{3}
\end{equation}
where $B$ is some known numerical constant. It is
worth noting that both results support the conjecture of the
exponential behavior of the multiparticle cross section\,(\ref{2})
with loop effects included.

The purpose of this letter is to demonstrate that the exponential
behavior \,(\ref{3}) of  the threshold  amplitudes is inherent in
all scalar theories (with constant $B$ being model-dependent)
and is not a feature  of $\lambda \phi ^4$ theory
only. We will see, in fact, that the technique developed
in refs.\,\cite{LRST,LST} can
be generalized to any scalar theory in a straightforward way.

\paragraph{2.} Let us consider a theory of one scalar field with the
action (we set the mass of the scalar boson  equal to one)
\begin{equation}
S=\int\!d^{d+1}x\left(\frac{(\partial _\mu \phi)^2 }{2}-
\frac{\phi ^2}{2}-V(\phi ,\{\lambda \})  \right)
\label{5}
\end{equation}
where $\{\lambda \}=\lambda _3,\,\,\lambda _4,\ldots $ is the set of
coupling constants and
\[
V(\phi ,\{\lambda \})=\sum\limits_{k=3}^{}\lambda _k\phi ^k
\]
In what follows we will assume that
the theory can be characterized by a unique weak coupling parameter
$\lambda _0$, so that the action is proportional to $\lambda_0
^{-1}$ after an appropriate rescaling.  This means that
\begin{equation}
\lambda _{k+1}
\sim\lambda _k^{\frac{k-1}{k-2}} \sim\lambda_0 ^{\frac{k-1}{2}}
\label{5*}
\end{equation}

Let us consider the process of creation of $n$ real bosons at
threshold (with $(d+1)$-momenta
equal to $(1,{\bf 0})$) by one virtual
particle with $(d+1)$-momentum $(n,{\bf 0})$ in the regime
$\{\lambda \}\to 0$, $\lambda _k^{\frac{2}{k-2}}n=$fixed,
$\lambda _k^{\frac{2}{k-2}}n\ll 1$.
This regime means, in particular, that we will be interested in
leading-$n$ behavior in each order of $\lambda_0 $.

As shown by Brown\,\cite{Brown}, the tree level amplitude of
the
process can be obtained by making use of the generating function,
\begin{equation}
A^{{\rm tree}}_{1\to n}=\frac{\partial ^n\phi _0(z(t))}{\partial z(t)^n}
\Big|_{z=0}
\label{6}
\end{equation}
where $\phi _0$ is the classical solution of the spatially
homogeneous (due to the threshold kinematics) field equation
\begin{equation}
\partial _t^2\phi _0+\phi _0+V'(\phi _0)=0
\label{6*}
\end{equation}
with the  boundary condition
\begin{equation}
\phi _0(t\to \infty)=z(t)+\ldots \equiv z_0{\rm e}^{it}+\ldots
\label{6**}
\end{equation}
where dots denote the terms suppressed by $\lambda _0$.
It is  convenient to introduce new euclidean time variable
\[
\tau =it+\ln z_0+C
\]
In the generic case
the constant $C$  can be chosen in such a way that $\phi _0$ is
singular at $\tau =0$,
\begin{equation} \phi _0(\tau \to 0)=\phi
_{\rm l}(\tau )+\ldots \label{6***} \end{equation}
where $\phi _{\rm l}(\tau )$ is the
leading singular term and dots mean less
singular ones.  One can easily find by scaling
that $\phi _{\rm l}\sim 1/\sqrt{\lambda _0}$. The leading
singularity of $\phi _0$ determines the leading-$n$ behavior of the
tree level amplitudes. Indeed, making use of the  Cauchy theorem one
finds
\begin{equation} A^{\rm tree}_{1\to n}=\frac{n!}{2\pi
i}\oint\!\frac{d\xi }{\xi ^{n+1}} \phi _0(\xi )\simeq
n!\int\limits_{C}^{2\pi i+C}\!d\tau {\rm e}^{-n\tau +nC}\phi _{\rm
l}(\tau )\sim n! \lambda _0^{\frac{n-1}{2}}a(n)
\label{7}
\end{equation}
where
$a(n)$ is a function that weakly depends on $n$.  For
example, $a\sim n^{m-1}$ for $\phi _{\rm l}\sim 1/\tau ^m$.  It is
worth noting that the tree level amplitude grows as $n!$ when
$n\to \infty$ in almost any
scalar theory (see, however, ref.~\cite{AKPU}).

Now let us concentrate on the loop corrections to the tree level
amplitude. In  complete analogy to the tree level case we will
use the generating function formalism. The full amplitude is given
by the following formula
\begin{equation} A^{{\rm
all-loop}}_{1\to n}=\frac{\partial ^n\langle\phi \rangle_{\phi
_0}}{\partial z(t)^n} \Big|_{z=0}
\label{8}
\end{equation}
where the
expectation value $\langle\phi \rangle $ is calculated in
the classical background $\phi _0$. So, extracting
the quantum part,
$\phi =\phi _0+\tilde{\phi }$, one can evaluate
$\langle\tilde{\phi }\rangle$ by  perturbation theory.
The corresponding Feynman rules
are presented in  Fig.\,\ref{Feyn} and diagram representation of
$\langle\tilde{\phi }\rangle$ is shown in  Fig.\,\ref{rep}.
\begin{figure}[tc]
\unitlength=1.00mm
\special{em:linewidth 0.4pt}
\linethickness{0.4pt}
\begin{picture}(120.00,15.00)(0,120)
\put(20.00,125.00){\line(1,0){18.00}}
\put(43.00,125.00){\makebox(0,0)[lc]{$=D(x,y)=\frac{1}{-\partial^2+1+V''(\phi_0)}$}}
\put(20.00,121.00){\makebox(0,0)[cc]{$x$}}
\put(38.00,121.00){\makebox(0,0)[cc]{$y$}}
\put(105.00,125.00){\circle*{1.00}}
\put(103.00,135.00){\line(1,-5){2.00}}
\put(105.00,125.00){\line(-1,-1){6.00}}
\put(105.00,125.00){\line(3,-4){5.33}}
\put(113.00,125.00){\line(-1,0){8.00}}
\put(105.00,125.00){\line(3,5){4.67}}
\put(120.00,125.00){\makebox(0,0)[lc]{$=-\frac{\partial^kV(\phi_0)}
{\partial\phi_0^k}$}}
\put(112.00,123.00){\makebox(0,0)[cc]{{\scriptsize 1}}}
\put(108.00,118.00){\makebox(0,0)[cc]{{\scriptsize 2}}}
\put(99.00,121.00){\makebox(0,0)[cc]{{\scriptsize 3}}}
\put(98.00,124.00){\circle*{0.6}}
\put(98.00,128.00){\circle*{0.60}}
\put(100.00,131.00){\circle*{0.60}}
\put(103.00,136.00){\makebox(0,0)[lc]{{\scriptsize $k-1$}}}
\put(111.00,132.00){\makebox(0,0)[cc]{{\scriptsize $k$}}}
\end{picture}
\caption{Feynman rules in the classical background $\phi _0$.}
\label{Feyn}
\end{figure}
\begin{figure}[hc]
\unitlength=1.00mm
\special{em:linewidth 0.4pt}
\linethickness{0.4pt}
\begin{picture}(129.00,30.00)(0,130)
\put(20.00,140.00){\makebox(0,0)[cc]{$\langle\tilde{\phi}
\rangle$\,\,\,\,=}}
\put(31.00,140.00){\line(1,0){14.00}}
\put(52.00,140.00){\circle{14.00}}
\put(64.00,140.00){\makebox(0,0)[cc]{+}}
\put(68.00,140.00){\line(1,0){10.00}}
\put(85.00,140.00){\circle{14.00}}
\put(82.00,140.00){\circle{8.00}}
\put(96.00,140.00){\makebox(0,0)[cc]{+}}
\put(100.00,140.00){\line(1,0){12.00}}
\put(119.00,140.00){\circle{14.00}}
\put(119.00,147.00){\line(0,-1){14.00}}
\put(129.00,140.00){\makebox(0,0)[lc]{$+\ldots$}}
\end{picture}
\caption{The diagram representation of $\langle\tilde{\phi
}\rangle$ in the background field $\phi _0$.}
\label{rep}
\end{figure}

Let us now study more closely the properties of the Euclidean
propagator
$D(x,x')$ in the classical background $\phi _0$. The propagator
satisfies the following equation ($\partial _0\equiv
\partial _\tau $),
\begin{equation}
(-\partial ^2_\mu +1+V''(\phi _0))D(x,x')=\delta ^{d+1}(x-x')
\label{9}
\end{equation}
and decays as $\tau \equiv x_0\to\pm\infty$. It is  convenient to
write the propagator in mixed coordinate-momentum representation
\[
D(x,x')=\int\!\frac{d^dp}{(2\pi )^d}{\rm e}^{i{\bf p(x-x')}}D_{\bf
p}(\tau ,\tau ')
\]
One  writes
\begin{equation}
D_{\bf p}(\tau ,\tau ')=\frac{1}{W_{\bf p}}(f_1^\omega(\tau )
f_2^\omega(\tau ') \theta (\tau' -\tau )+ f_2^\omega(\tau )
f_1^\omega(\tau ') \theta (\tau -\tau '))
\label{9*}
\end{equation}
where $f_1^\omega(\tau )$ and $f_2^\omega(\tau )$ are two linearly
independent solutions to the homogeneous  equation
\begin{equation}
(-\partial ^2_\tau +\omega ^2+V''(\phi _0))f(\tau )=0
\label{9**}
\end{equation}
with $\omega =\sqrt{{\bf p^2}+1}$. $f_1^\omega(\tau )$ and
$f_2^\omega(\tau )$ tend to zero as $\tau \to-\infty$ and $\tau
\to\infty$, respectively. Note, that $\phi _0(\tau \to\pm\infty)\to 0$,
$V''(\phi _0(\tau \to\pm\infty))\to 0$, and,
therefore, $f_1^\omega (\tau
\to-\infty)\to{\rm e}^{\omega \tau }$, $f_2^\omega (\tau
\to\infty)\to{\rm e}^{-\omega \tau }$.
Finally in eq.\,(\ref{9*}) we introduced the notation
\begin{equation}
W_{\bf p}=f_1'f_2-f_2'f_1
\label{10+}
\end{equation}
$W_{\bf p}$ is the Wronskian
which does not depend on $\tau $.

Let us now concentrate on the behavior of $f_1$, $f_2$, and $D_{\bf
p}(\tau ,\tau ')$ at $\tau \to 0$. One  first notes that
$V''(\phi _0(\tau \to 0))=\phi _{\rm l}'''/
\phi _{\rm l}'$ due to the field equation \,(\ref{6*}).
Generically, both $f_1$ and $f_2$ are singular at $\tau =0$, and
one can normalize them in such a way that
\begin{equation}
f_1(\tau )=f_2(\tau )=\phi _{\rm l}'(\tau )+\ldots
\label{10*}
\end{equation}
in the sense of the leading
 singularity at $\tau \to 0$. However, equation\,(\ref{10*})
means that $f_1$ and $f_2$ are linearly
dependent at $\tau \to0$. Therefore,
it is  convenient to introduce the new basis
\[
f=\frac{f_1+f_2}{2};\,\,\,\,g=\frac{f_1-f_2}{2}
\]
These functions have different behavior near $\tau =0$: while $f$ is
most singular, $f(\tau \to0)=\phi' _{\rm l}(\tau )$, the function
$g$ is less singular or regular at this point. To study the
behavior of $g$ at $\tau =0$ one makes use of the independence of
$W_{\bf p}$ on $\tau $.  From this condition one obtains
\begin{equation}
g(\tau \to0)=\frac{W_{\bf p}}{2}\phi _{\rm l}(\tau
)'\int\limits_{0}^{\tau} \!\frac{d\xi }{(\phi _{\rm l}(\xi )')^2}
\label{g}
\end{equation}
In terms of $f$ and $g$, the propagator can be represented in the
following form,
\[
D(\tau ,\tau ';{\bf p})=D_0(\tau ,\tau ';{\bf p}) +
D_1(\tau ,\tau ';{\bf p})
\]
where
\[
D_0(\tau ,\tau ';{\bf p})=\frac{1}{W_{\bf p}}f(\tau )f(\tau ')
\]
\[
D_1(\tau ,\tau ';{\bf p})=\frac{1}{W_{\bf p}}[\epsilon (\tau -\tau
')(f(\tau ) g(\tau ')-f(\tau ')g(\tau ))-g(\tau )g(\tau ')] \]
$D_0$ contains the strongest singularity of $D$, while $D_1$ is less
singular than $D_0$. Note that $D_0(\tau ,\tau ')$
factorizes
\[
D_0(\tau ,\tau ')\simeq\frac{1}{W_{\bf p}}
\phi _{\rm l}(\tau )'\phi _{\rm l}(\tau ')'
\]
while the leading term in
$D_1$ at $\tau ,\,\tau '\to0$ does not depend on ${\bf p}$, see eq.\,
(\ref{g}).
These two properties show remarkable similarity to the
$\lambda \phi ^4$-theory\,\cite{LRST} and, as in the $\lambda
\phi ^4$-theory, will be extensively used in what follows.

Now we are ready to proceed to the analysis of the loop
corrections. At
first sight, to find the leading singular term in each loop
one has to replace all propagators $D$ by $D_0$. This is, however, not
correct. The reason is that $D_0(\tau ,\tau ')$ is smooth at $\tau
=\tau '$, so that the contribution proportional, for instance, to
$D_0^2$ is less singular than the contribution that comes from the
product $D_0D_1$. This is precisely the argument of
ref.~\cite{LRST}. It is straightforward to check that the leading
singular term in each loop  can be obtained by replacement of $D$
by $D_0$ and $D_1$ in the following way. First, the obtained graphs
should not contain closed loops consisting of $D_1$ only. Second,
these graphs should not factorize, in spite of the
factorization of $D_0$.  In particular, the number of
$D_0$'s should coincide with the  number of loops.

After this procedure, the integration over internal loop momenta
becomes trivial because of the fact that the entire dependence on ${\bf
p}_1,\ldots,{\bf p}_k$ ($k$ is the number of loops) is through the
product $\prod\limits_{i=1}^{k}W_{{\bf p}_i}$.  So, after
integrating over momenta one finds that the contribution from
the $k$-th loop is proportional to $b^k$, where
\begin{equation}
b=\frac{1}{2}\int\limits_{}^{}\frac{d^dp}{(2\pi )^d}\frac{1}{W_{\bf
p}}
\label{14*}
\end{equation}
By a simple counting argument one
can see that $b\sim\lambda _0$, $ b=
\lambda _0B$. Furthermore, it
follows from eqs.\,(\ref{9**}), \,(\ref{10+}),\,(\ref{10*}) that
the Wronskian grows as $\phi _{\rm l}(\tau )'' \phi _{\rm l}(\tau
)'|_{\tau =1/\omega }$ when $\omega $ tends to infinity.  So, at
$d\le 3$  the integral
\,(\ref{14*}) is generically convergent in the ultraviolet, with an
exception of the case $d=3$ and
$\lim\limits_{\tau \to0}\phi _{\rm l}\tau ^m=0$ for any $m>0$.

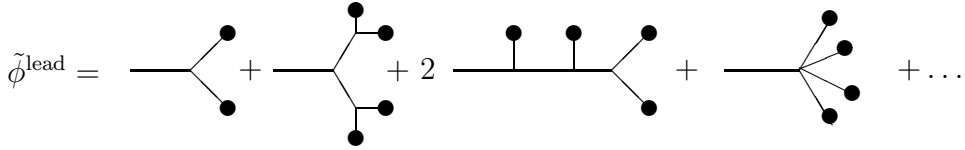
\begin{figure}[tc] \unitlength=1mm
\special{em:linewidth 0.4pt}
\linethickness{0.4pt}
\begin{picture}(129.00,15.00)(0,128)
\put(23.00,135.00){\makebox(0,0)[rc]{$\tilde{\phi}^{\rm lead}=$}}
\put(27.00,135.00){\line(1,0){8.00}}
\put(35.00,135.00){\line(1,1){5.00}}
\put(35.00,135.00){\line(1,-1){5.00}}
\put(40.00,130.00){\circle*{2.00}}
\put(40.00,140.00){\circle*{2.00}}
\put(43.00,135.00){\makebox(0,0)[cc]{+}}
\put(46.00,135.00){\line(1,0){8.00}}
\put(54.00,135.00){\line(3,5){3.00}}
\put(57.00,140.00){\line(0,1){4.00}}
\put(57.00,140.00){\line(1,0){4.00}}
\put(61.00,140.00){\line(0,0){0.00}}
\put(54.00,135.00){\line(3,-5){3.00}}
\put(57.00,130.00){\line(1,0){4.00}}
\put(57.00,130.00){\line(0,-1){4.00}}
\put(57.00,126.00){\circle*{2.00}}
\put(61.00,130.00){\circle*{2.00}}
\put(61.00,140.00){\circle*{2.00}}
\put(57.00,143.00){\circle*{2.00}}
\put(65.00,135.00){\makebox(0,0)[cc]{+ 2 }}
\put(70.00,135.00){\line(1,0){21.00}}
\put(91.00,135.00){\line(1,1){5.00}}
\put(91.00,135.00){\line(1,-1){5.00}}
\put(96.00,130.00){\line(0,0){0.00}}
\put(86.00,135.00){\line(0,1){5.00}}
\put(78.00,140.00){\line(0,-1){5.00}}
\put(96.00,130.00){\circle*{2.00}}
\put(96.00,140.00){\circle*{2.00}}
\put(86.00,140.00){\circle*{2.00}}
\put(78.00,140.00){\circle*{2.00}}
\put(101.00,135.00){\makebox(0,0)[cc]{+}}
\put(106.00,135.00){\line(1,0){10.00}}
\put(116.00,135.00){\line(3,5){4.33}}
\put(123.00,138.00){\line(-5,-2){7.00}}
\put(116.00,135.33){\line(2,-1){7.00}}
\put(116.00,135.00){\line(3,-5){4.33}}
\put(120.00,129.00){\circle*{2.00}}
\put(123.00,132.00){\circle*{2.00}}
\put(122.00,138.00){\circle*{2.00}}
\put(120.00,142.00){\circle*{2.00}}
\put(129.00,135.00){\makebox(0,0)[lc]{$+\ldots$}}
\end{picture}
\caption{Graphs contributing to the leading singularity of
$\tilde{\phi} $.}
\label{tree}
\end{figure}

The procedure described above  can be represented graphically. In
Fig.\,\ref{tree} several so called bullet~\cite{LRST} graphs are
shown.  These diagrams correspond to ones presented in
Fig.\,\ref{rep}.
Let us
explain some details concerning Fig.\,\ref{tree}. First,
each line that ends at a bullet corresponds to a factor
$(2b)^{1/2}\phi _{\rm l}'$. Second, each pair of bullets
corresponds to an internal
line in Fig.\,\ref{rep} which have been replaced by $D_0$. Because of
the factorization of
$D_0$, the internal line can be cut and reduced to two bullets.
Third, all
bullet diagrams are tree and connected. Therefore the problem of
calculation of loop contributions reduces to the summation of
certain tree graphs.  To perform this calculation, we note that the
bullet diagrams have the same form as the graphs in an effective
theory where the condensate $\phi _0$ is shifted by $(2b)^{1/2}\phi
_{\rm l}'$, with the only difference in the symmetry factor:  each
graph with $2k$ external lines ending at bullets must be multiplied
by a factor of $(2k)!/(2^kk!)$. So, we proceed follows.
We search for a solution to the classical field equation (which
is equivalent to the summation of tree graphs)
near the singularity,
 \begin{equation}
\partial _\tau ^2\phi_{\rm cl} -V'(\phi_{\rm cl} )=0
\label{16}
\end{equation} (the
mass can be neglected) which at small $\lambda _0$
has the form \begin{equation} \phi
_{\rm cl}=\phi _{\rm l}+\sqrt{2b}\phi _{\rm l}'+
O(\lambda _0)
\label{16*}
\end{equation} The solutions of eq.\,(\ref{16}) are known,
\begin{equation}
\phi _{\rm cl}=\phi _{\rm l}(\tau +\alpha )
\label{16**}
\end{equation}
again in the sense of the leading singularity.
Comparing eqs.\,(\ref{16*}) and\,(\ref{16**}) we obtain $\alpha
=\sqrt{2b}$ and, therefore,
\begin{equation}
\phi _{\rm cl}=\sum\limits_{i=0}^{\infty}\frac{(2b)^{\frac{i}{2}}}{i!}
\frac{\partial ^i\phi _{\rm l}(\tau )}{\partial \tau ^i}
\label{16***}
\end{equation}
The $i$-th term of this series corresponds to the sum of tree graphs
ending at $i$ bullets. To recover the generating function $\langle
\phi \rangle$ one should omit all terms in eq.\,(\ref{16***}) with odd
$i$ and multiply  each term with even $i$, $i=2k$,
by the factor $(2k)!/(2^kk!)$. In this way we obtain
\begin{equation}
\langle\phi \rangle_{\phi _0}=\sum\limits_{k=0}^{\infty}
\frac{b^k}{k!}
\frac{\partial ^{2k}\phi _{\rm l}(\tau )}{\partial \tau ^{2k}}
\label{17}
\end{equation}
Substituting this formula into eq.\,(\ref{8}), differentiating by means
of the Cauchy theorem, and using eq.\,(\ref{7}) we finally get
\begin{equation}
A_{{\rm all-loop}}(\epsilon =0)=A_{{\rm tree}}{\rm e}^{b
n^2}
\label{17*}
\end{equation}
Recall that $b$ is of order  $\lambda _0$.

Therefore, we have shown  that the
exponentiation of loop corrections is inherent in all scalar
theories.  The  technique    can be
straightforwardly generalized to the case of the initial
state with
several particles\,\cite{LST} and to scalar theories with several
fields.  However, in exceptional models it may happen that
  the basic assumption that $\phi
_0(\tau )$ is singular at $\tau =0$ (or, in other words,
the  singularity of $\phi _0$ is not at
infinity) is not correct. For example, in the theory of
ref.\,\cite{AKPU} with $V(\phi )=(1+\lambda \phi
)^2\ln^2(1+\lambda \phi )^2/(8\lambda ^2)-\phi ^2/2$ one finds
\cite{AKPU}
\[
\phi _0(t)=\frac{1}{\lambda }\left({\rm e}^{\lambda z(t)}-1 \right)
\]
so that $A_{1\to n}^{\rm tree}=\lambda ^{n-1}$ -- the tree level
amplitudes do not grow
factorially! The reason, of course, is that the solution $\phi _0$
has a singularity only at  $\tau \equiv it+\ln\lambda
z_0\to\infty$.  Moreover, more singular terms, for instance
$\phi _0^2$ or $\phi _0'$, contribute to the amplitude
in the same order as $\phi
_0$, so the technique described in this letter can not be applied to
theories of this type.

\section*{Acknowledgments}

\noindent
The author is indebted to A.Yu.~Morozov, V.A.~Rubakov, and
S.V.~Troitsky for helpful discussions. The work is supported in
part by Russian Foundation for Basic Research grant 96-02-17449a,
INTAS grant 94-2352, and by Soros fellowship for
graduate students.

\end{document}